\documentclass[a4paper,11pt]{article}
\usepackage{pos}
\usepackage{multirow}
\usepackage{multicol}
\usepackage{tabu}
\usepackage{array}
\newcolumntype{C}[1]{>{\centering\let\newline\\\arraybackslash\hspace{0pt}}m{#1}}
\usepackage{graphicx}
\usepackage{wrapfig}
\newlength{\bibitemsep}
\newlength{\bibparskip}
\let\oldthebibliography\thebibliography
\renewcommand\thebibliography[1]{%
  \oldthebibliography{#1}%
  \setlength{\parskip}{\bibitemsep}%
  \setlength{\itemsep}{\bibparskip}%
}

\title{Neutrino Target-of-Opportunity Observations with Space-based and Suborbital Optical Cherenkov Detectors}
 \ShortTitle{Neutrino ToO Observations with Space and Suborbital Cherenkov Detectors}

\author*[a]{Tonia M. Venters}
\author[b]{Mary Hall Reno}
\author[a,c,d]{John F. Krizmanic}
\forColls{POEMMA and JEM-EUSO} 

\affiliation[a]{NASA Goddard Space Flight Center,\\
  Astrophysics Science Division, Greenbelt, Maryland, USA}

\affiliation[b]{University of Iowa,\\
  Department of Physics and Astronomy, Iowa City, Iowa, USA}
  
\affiliation[c]{University of Maryland, Baltimore County,\\ Center for Space Sciences and Technology,\\
  Baltimore, Maryland, USA}

\affiliation[d]{Center for Research and Exploration in Space Science \& Technology}

\emailAdd{tonia.m.venters@nasa.gov}
\emailAdd{mary-hall-reno@uiowa.edu}

\abstract{Cosmic-ray accelerators capable of reaching ultra-high energies are expected to also produce very-high energy neutrinos via hadronic interactions within the source or its surrounding environment. Many of the candidate astrophysical source classes are either transient in nature or exhibit flaring activity. Using the Earth as a neutrino converter, suborbital and space-based optical Cherenkov detectors, such as POEMMA and EUSO-SPB2, will be able to detect upward-moving extensive air showers induced by decaying tau-leptons generated from cosmic tau neutrinos with energies $\sim 10$ PeV and above. Both EUSO-SPB2 and POEMMA will be able to quickly repoint, enabling rapid response to astrophysical transient events. We calculate the transient sensitivity and sky coverage for both EUSO-SPB2 and  POEMMA, accounting for constraints imposed by the Sun and the Moon on the observation time. We also calculate both detectors' neutrino horizons for a variety of modeled astrophysical neutrino fluences. We find that both EUSO-SPB2 and POEMMA will achieve transient sensitivities at the level of modeled neutrino fluences for nearby sources. We conclude with a discussion of the prospects of each mission detecting at least one transient event for various modeled astrophysical neutrino sources.}

\FullConference{37$^{\rm{th}}$ International Cosmic Ray Conference (ICRC 2021)\\
		July 12th -- 23rd, 2021\\
		Online -- Berlin, Germany}

\begin{document}
\maketitle

\section{Introduction}
\vspace{-2.0ex}

Transient phenomena underpinned both of the recent landmark multimessenger events--the coincident neutrino and electromagnetic detection of a flare in blazar TXS0506+056~\cite{IceCube:2018dnn} and the coincident gravitational wave and electromagnetic detection of GW170817, a binary neutron star merger accompanied by a short gamma-ray burst~\cite{TheLIGOScientific:2017qsa,Monitor:2017mdv}. These events solidified the central role of Target-of-Opportunity (ToO) observations in the emerging multimessenger landscape for the coming era and beyond. 

Many astrophysical transient sources are suspected of producing high-energy and very-high energy neutrinos, including tidal disruption events~\cite[\textit{e.g.},][]{Lunardini:2016xwi}, compact object mergers~\cite[\textit{e.g.},][]{Kotera:2016dmp,Fang_BNNMerger}, gamma-ray bursts~\cite[\textit{e.g.},][]{Waxman:1997ti,Kimura:2017kan}, blazar flares~\cite[\textit{e.g.},][]{Waxman:1998yy}, and possibly others. In typical astrophysical models, neutrinos\footnote{In this work, we do not distinguish between neutrinos and antineutrinos.} 
are primarily produced through the decay of charged pions and secondary muons generated through interactions of accelerated cosmic rays with matter and radiation within the source~\cite[\textit{e.g.},][]{Lipari:2007su}. The expected at-source flavor ratio from pion decay $(\nu_e : \nu_{\mu} : \nu_{\tau})_{\rm src}$, is nearly $(1 : 2 : 0)$; though, neutrino oscillations drive the observed flavor ratio towards $(1 : 1 : 1)_{\bigoplus}$. 
This provides a channel through which indirect cosmic ray experiments may observe transient astrophysical neutrino sources by detecting upward-going extensive air showers (EASs) induced by decaying $\tau$-leptons produced by $\nu_{\tau}$ interactions in the Earth.

In this work, we focus on the unique contributions to multimessenger studies of astrophysical transients from the proposed space-based Probe of Extreme MultiMessenger Astrophysics (POEMMA)~\cite{POEMMA-JCAP} and the upcoming second-generation Extreme Universe Space Observatory on a Super-Pressure Balloon (EUSO-SPB2)~\cite{EUSO_SPB2_ICRC21}. Both missions will monitor the Earth's atmosphere in search of fluorescence\footnote{The fluorescence detection capabilities of POEMMA and EUSO-SPB2 are discussed in other proceedings.} signals from ultra-high energy cosmic rays and neutrinos above $20$~EeV and optical Cherenkov signals from upward-going EASs of tau neutrinos above $10$~PeV. In order to detect these signals, both missions must operate during dark, moonless night conditions. While it is possible that POEMMA and EUSO-SPB2 could generate their own transient alerts, this work focuses on their capabilities for following up alerts generated by other multimessenger facilities. Table~\ref{tab:Miss_Specs} provides a select list of relevant observatory specifications for POEMMA and EUSO-SPB2. 

The POEMMA observatory design consists of two telescopes on board individual satellites orbiting the Earth in tandem with a period of $\sim 95$~min. Each spacecraft will be equipped with avionics that will allow them to quickly repoint (at a rate of $90^{\circ}$ in $\sim 500$~s) in the direction of a transient source and track it through the tau neutrino detection region (up to $18^{\circ}$ below the Earth's limb). With these design features, POEMMA will have access to the entire dark sky within the time scale of one orbit and full-sky coverage within a few months. POEMMA is proposed for launch in 2028 and has a target mission lifetime of $3$--$5$~years.

EUSO-SPB2 is a suborbital payload experiment that will be a pathfinder for POEMMA in many respects. It will feature a similar, though simplified, optical design for validating the technology for detecting optical Cherenkov signals from Earth-skimming tau neutrinos. It will also have a ToO operation mode that will enable pointing to a transient source by tilting the telescope in elevation and rotating the entire gondola in azimuth. The expected launch for EUSO-SPB2 is in Spring 2023 from Wanaka, New Zealand, with a target flight duration of $\sim 100$~days.
\begin{table}[t!]
    \centering
    \begin{tabular}{|C{1.5in}|C{1.25in}|C{1.25in}|}
    \hline
    \hline
        Observatory Specification & POEMMA & EUSO-SPB2 \\
        \hline
        \hline
        Altitude ($h$) & $525$~km & $33$~km \\ 
        \hline
        Field-of-View ($\Delta \phi \times \Delta \alpha$) & $30^{\circ} \times 9^{\circ}$ & $12.8^{\circ} \times 6.4^{\circ}$ \\
        \hline
        Max. Angle from Limb & $18^{\circ}$ & $6.4^{\circ}$ \\
        \hline
        Detection threshold & $20$ ($40$) phot m$^{-2}$ & $200$ phot m$^{-2}$ \\
        \hline
        Max. sky coverage & 100\% & 74\% \\
        \hline
        \hline
    \end{tabular}
    \caption{\textit{Select list of observatory specifications for POEMMA and EUSO-SPB2.}} 
    \label{tab:Miss_Specs}
    \vspace{-1.0ex}
\end{table}

\section{\texorpdfstring{$\mathbf{\nu_{\tau}}$}{nt} Acceptance and Transient Sensitivity Including Sun and Moon Constraints}\label{sec:AccAndSens}
\vspace{-2.0ex}

\begin{wrapfigure}{r}{0.5\textwidth}
    \vspace{-4.0ex}
    \centering
    \includegraphics[width=.48\textwidth, trim = 0 1.0in 4.75in 3.0in, clip]{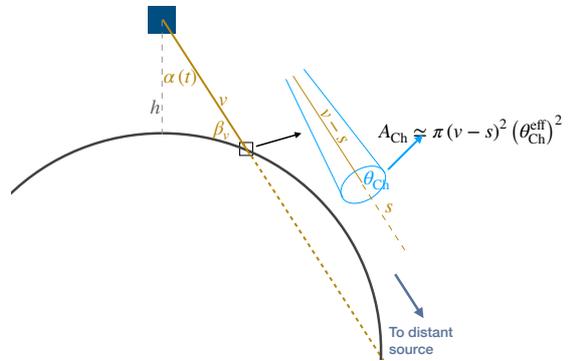}
    \caption{\textit{$\nu_{\tau}$ detection geometry for point sources.}}
    \label{fig:nuTauDetGeom}
    \vspace{-8.0ex}
\end{wrapfigure}

The $\nu_{\tau}$ acceptance and the transient sensitivity depend on the instantaneous effective area to point sources (see Fig.~\ref{fig:nuTauDetGeom}), which is defined as the area subtended on the ground by the Cherenkov cone given by~\cite{ToO_poemma,HallsieNuTau},
\begin{equation}
    A_{\rm Ch} \simeq \pi (v-s)^2 \times \left( \theta^{\rm eff}_{\rm Ch}\right)^2\,,
\end{equation}
where $v$ is the distance between the detector at altitude $h$ and the spot on the ground along the line-of-sight to a point source in the $\nu_{\tau}$ detection region, $s$ is the decay length for the $\tau$-lepton, and $\theta^{\rm eff}_{\rm Ch}$ is the effective Cherenkov angle for upward-going EASs. The line-of-sight is viewed at an angle $\alpha(t)$ from detector nadir, corresponding to an elevation angle $\beta_{v}$ with respect to the spot on the ground. The effective Cherenkov angle depends on the altitude at which the $\tau$-lepton decays, the shower energy, and elevation angle 
of the $\tau$-lepton emerging from the ground, $\beta_{\rm tr}$~\cite{Reno:2019jtr}. 
Since the neutrino sources in question are distant 
point sources, we take $\beta_{\rm tr} \simeq \beta_{v}$.
The instantaneous acceptance to tau neutrinos with energy $E_{\nu_{\tau}}$ is given by
\begin{equation}
    A\left(\alpha(t),E_{\nu_{\tau}}\right) \simeq \int dP_{\rm obs}\left(E_{\nu_{\tau}},\beta_{v},s\right)A_{\rm Ch}(s)\,,
\end{equation}
where $P_{\rm obs}$ is the observation probability. $P_{\rm obs}$ is related to the probability for a $\tau$-lepton to emerge at angle $\beta_{v}$ (\textit{i.e.}, the tau exit probability, $p_{\rm exit}$), the $\tau$-lepton decay distribution
($p_{\rm dec}$), and the detection 
probability ($p_{\rm det}$) by
\begin{equation}
    P_{\rm obs} = \int \!\!\! \int p_{\rm exit}\left(E_{\tau}|E_{\nu_{\tau}},\beta_{v}\right) p_{\rm dec}(s') p_{\rm det}\left(E_{\tau},\theta^{\rm eff}_{\rm Ch},\beta_{v},s'\right) ds' dE_{\tau}\,.
\end{equation}
%
%
For $p_{\rm exit}$, we used the results from~\cite{Reno:2019jtr}, which performed detailed simulations of $\nu_{\tau}$ propagation through the Earth~\cite[see also][]{nuPyProp}.
\cite{Reno:2019jtr} also performed detailed simulations propagating Cherenkov signals from upward-going EASs through the atmosphere to a space-based detector flying at altitude $h = 525$~km; we used these results to determine $\theta^{\rm eff}_{\rm Ch}$ and $p_{\rm det}$ for POEMMA. We performed new Cherenkov simulations to determine $\theta^{\rm eff}_{\rm Ch}$ and $p_{\rm det}$ for EUSO-SPB2~\cite[see also][]{Cummings:2020ycz}.

As the source moves through the detection region, the effective area and the $\nu_{\tau}$ acceptance change with time. In determining the transient sensitivity, we compute the time-averaged $\nu_{\tau}$ acceptance,
\begin{equation}
    \left<A\left(E_{\nu_{\tau}}\right)\right> = \frac{1}{T_{0}} \int^{t_0 + T_{0}}_{T_{0}} dt A\left(\alpha(t),E_{\nu_{\tau}}\right)\,.
\end{equation}
The all-flavor transient sensitivity is then given by
\begin{equation}
    {\rm Sensitivity} = \frac{2.44 \times 3}{\ln(10) \times E_{\nu_{\tau}} \times \left<A\left(E_{\nu_{\tau}}\right)\right>}\,,
\end{equation}
where we have taken the $90$\% unified confidence upper limit~\cite{Feldman:1997qc} per decade in energy ($2.44$/$\ln(10)$). In all scenarios considered here, we assume that both instruments can slew over $360^{\circ}$ in azimuth.

In this work, we calculate separate estimates for the transient sensitivity for long-duration and short-duration events. For short-duration events, the relevant time scale for computing the time-averaged $\nu_{\tau}$ acceptance is the observation time, which we take to be $\sim 10^3$~s. For long-duration events (those lasting $\gtrsim 1$~day), the relevant time scale 
is one that allows for adequate sampling of the variation in the effective area as the source moves through the detection region. For POEMMA, this time scale is $T_0 \simeq T_{\rm orb} = 95$~minutes. For EUSO-SPB2, we set $T_0$ equal to the duration of the balloon flight in order to more robustly account for the constraints due to the Sun and the Moon (as discussed below). In this work, we consider times of $30$~days and $100$~days for a typical balloon flight and the target flight duration, respectively.

During long-duration events (time scales $\gtrsim 1$~day), POEMMA's spacecraft propulsion systems can maneuver its satellites closer together (separation $\sim 25$~km) in order to observe the source within overlapping instrument light pools (denoted ToO-stereo configuration). This allows for the use of time coincidence between the two instruments, enabling better rejection of the night-sky air glow background and lowering the energy threshold for neutrino detection. For events shorter than a day, the two POEMMA telescopes will observe the source independently in separate light pools (denoted ToO-dual configuration); this configuration doubles the effective area at the expense of observing with a higher energy threshold.

Both POEMMA and EUSO-SPB2 must operate in dark, moonless night conditions. We account for the effects of the Sun and the Moon depending on the scenario under consideration. For long-duration events, the time scales are sufficiently long that the Sun and the Moon are guaranteed to constrain the observation period. For space-based instruments, the reduction in exposure depends on the sky location of the source and the relative positions of the Sun, the Moon, the Earth, and position of the satellite in its orbit, resulting in a large number of possible configurations. For POEMMA, the average duty cycle\footnote{Averaged over seven precession periods of POEMMA's orbital plane ($7 \times 54.3$~days $\simeq 380$~days).} ranges between 0.2--0.4 over the sky~\cite{ToO_poemma}~\cite[see also][]{Guepin:2018yuf}; for our calculations of the POEMMA's long-duration transient sensitivity, we adopt a value of $30$\%. 

\begin{figure}
    \centering
    \includegraphics[width=0.49\textwidth, trim = 1.75in 2.0in 1.75in 2.35in, clip]{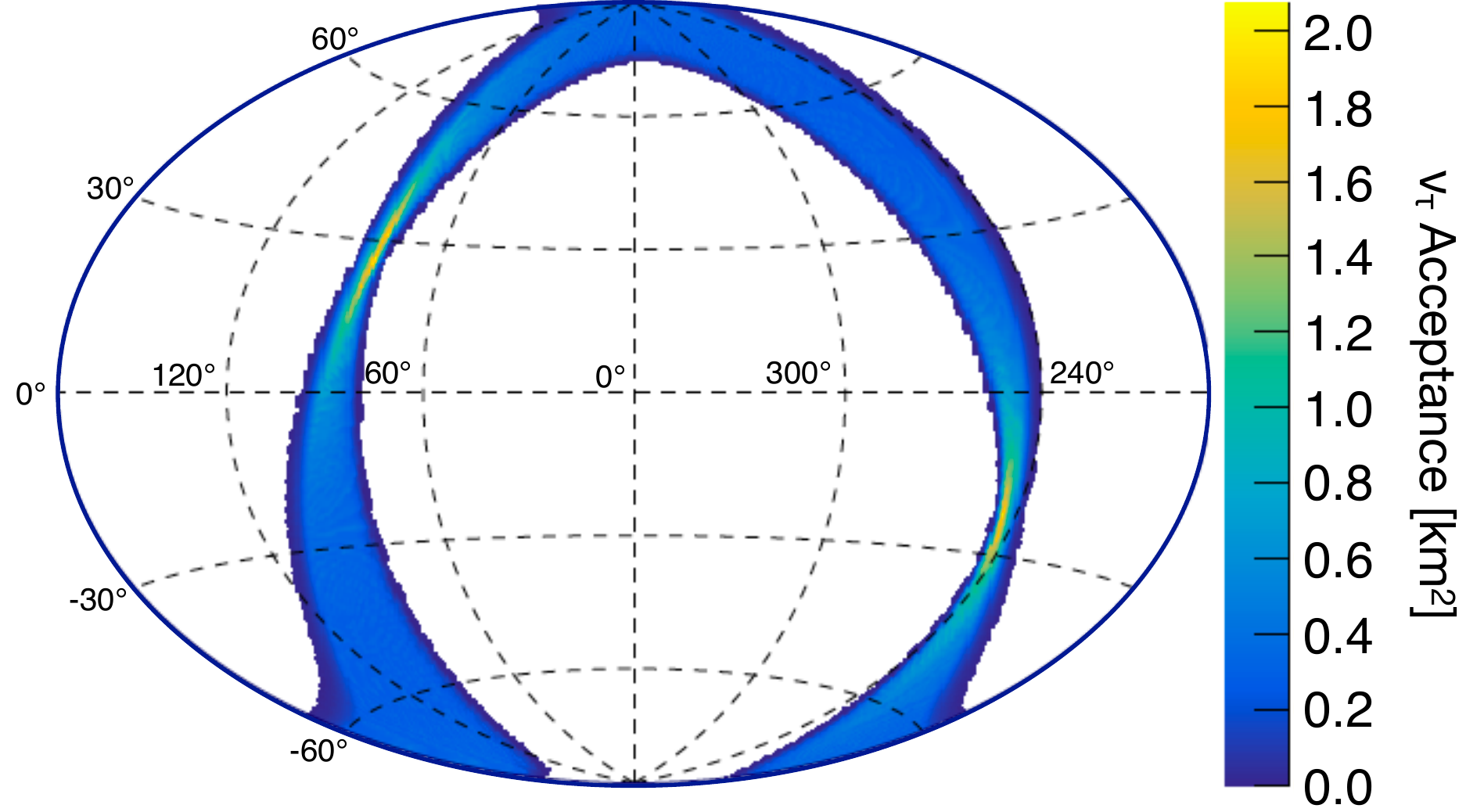}\includegraphics[width=0.49\textwidth, trim = 1.75in 2.0in 1.75in 2.35in, clip]{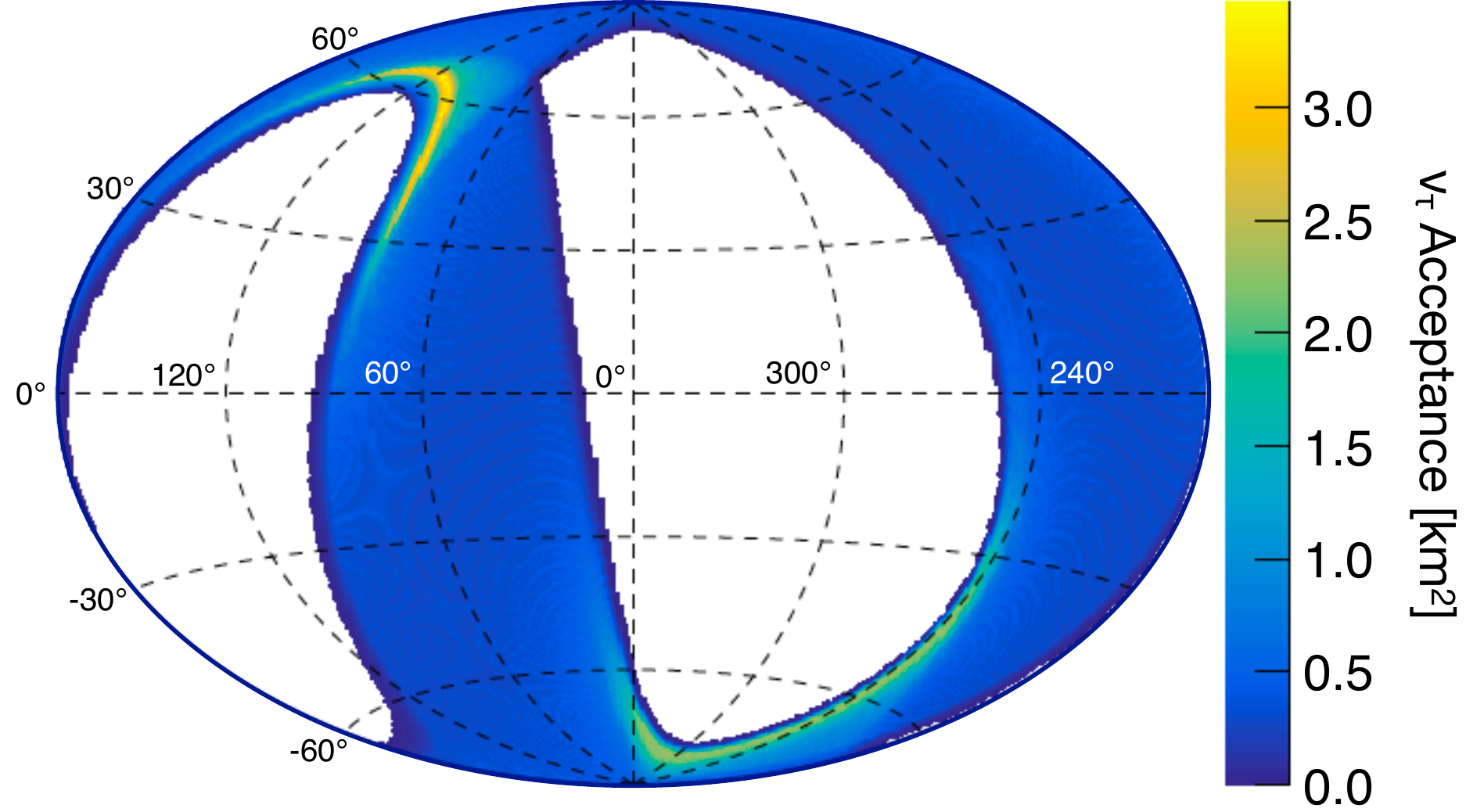}
    \caption{\textit{Sky plots of the $\nu_{\tau}$ acceptance at $10^{8.5}$~GeV averaged over a period of a day when the Moon is close to full (left) and close to new (right).}}
    \label{fig:SPB2SunMoonSkyPlots}
    \vspace{-3.0ex}
\end{figure}

For EUSO-SPB2, we perform a more detailed calculation of the average $\nu_{\tau}$ acceptance, accounting for the reduction in observation time due to the Sun and the Moon. For the Sun, we set the acceptance equal to zero during those time periods in which the elevation of the Sun was greater than $18^{\circ}$ below the horizon as viewed by EUSO-SPB2 (elevation angle of $-5.8^{\circ}$). For the Moon, we calculate the average intensity of moonlight, including direct and scattered components, accounting for the phase. To that end, we used libRadtran, a radiative transfer library~\cite{2016GMD.....9.1647E}, 
to calculate the average intensity of sunlight ($I_{\bigodot}$) 
for an instrument flying at $33$~km, assuming 
an Earth albedo of $\sim 0.4$~\cite{Peyrou_Lauga_Earth_Albedo} and an atmospheric profile consistent with midlatitude winter and accounting for aerosols. The moonlight intensity is found by rescaling the solar intensity,
%
%
\begin{equation}
    I_{\rm Moon} = 0.07 I_{\bigodot} \left(\frac{R_{\rm Moon}}{d_{\bigoplus - {\rm Moon}}}\right)^2 10^{-0.4\left(0.026|i| + 4.0 \times 10^{-9}i^4\right)}\,,
\end{equation}
where $R_{\rm Moon}$ is the radius of the Moon, $d_{\bigoplus - {\rm Moon}}$ is the distance between the Earth and the Moon, and the last multiplicative factor accounts for the variation in intensity due to the phase of the Moon with $i$ being the lunar phase angle in degrees~\cite{1991PASP..103.1033K}.
\footnote{Note that $0^{\circ} \leq i \leq 180^{\circ}$, where $i = 0^{\circ}$ corresponds to a full moon and $i = 180^{\circ}$ corresponds to a new moon.} For the constraint due to the Moon, we set the acceptance equal to zero during those time periods in which $I_{\rm Moon}$ is greater than $10$\% of the estimated intensity of the night sky air glow ($\sim 18801$ phot m$^{-2}$ ns$^{-1}$ sr$^{-1}$ integrated from $300$--$1000$~nm to cover the wavelength range of the optical Cherenkov telescope). The impact of the Sun and the Moon can be seen in Figure~\ref{fig:SPB2SunMoonSkyPlots}, which plots the $\nu_{\tau}$ acceptance averaged over a period of a day for two different phases of the Moon.

For short-duration events for both missions, the time scale of the observation considered here ($10^3$~s) is sufficiently short that observations may occur without interference from the Sun or the Moon. In this work, we consider short-duration scenarios in which observations are not constrained by the Sun and the Moon. We also assume that event occurs when the source is in the optimal position for $\nu_{\tau}$ detection (\textit{i.e.}, when source just dips below the Earth's limb). As such, our short-duration scenarios should be considered ``best-case'' scenarios for neutrino ToO observations.

\begin{figure}
    \centering
    \raisebox{-0.5\height}{\includegraphics[width=.45\textwidth]{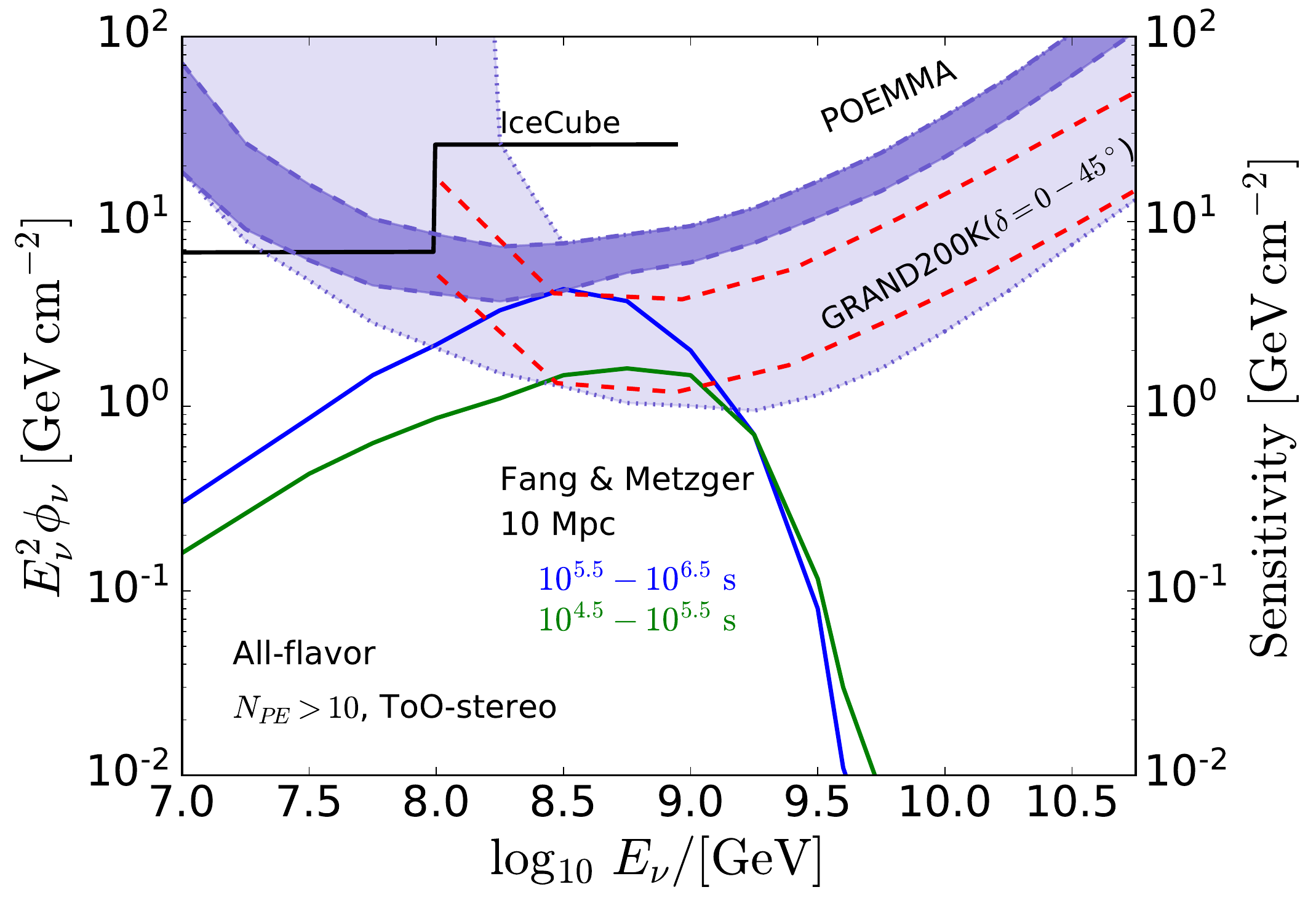}}\raisebox{-0.5\height}{\includegraphics[width=0.49\textwidth, trim = 1.75in 2.0in 1.75in 2.35in, clip]{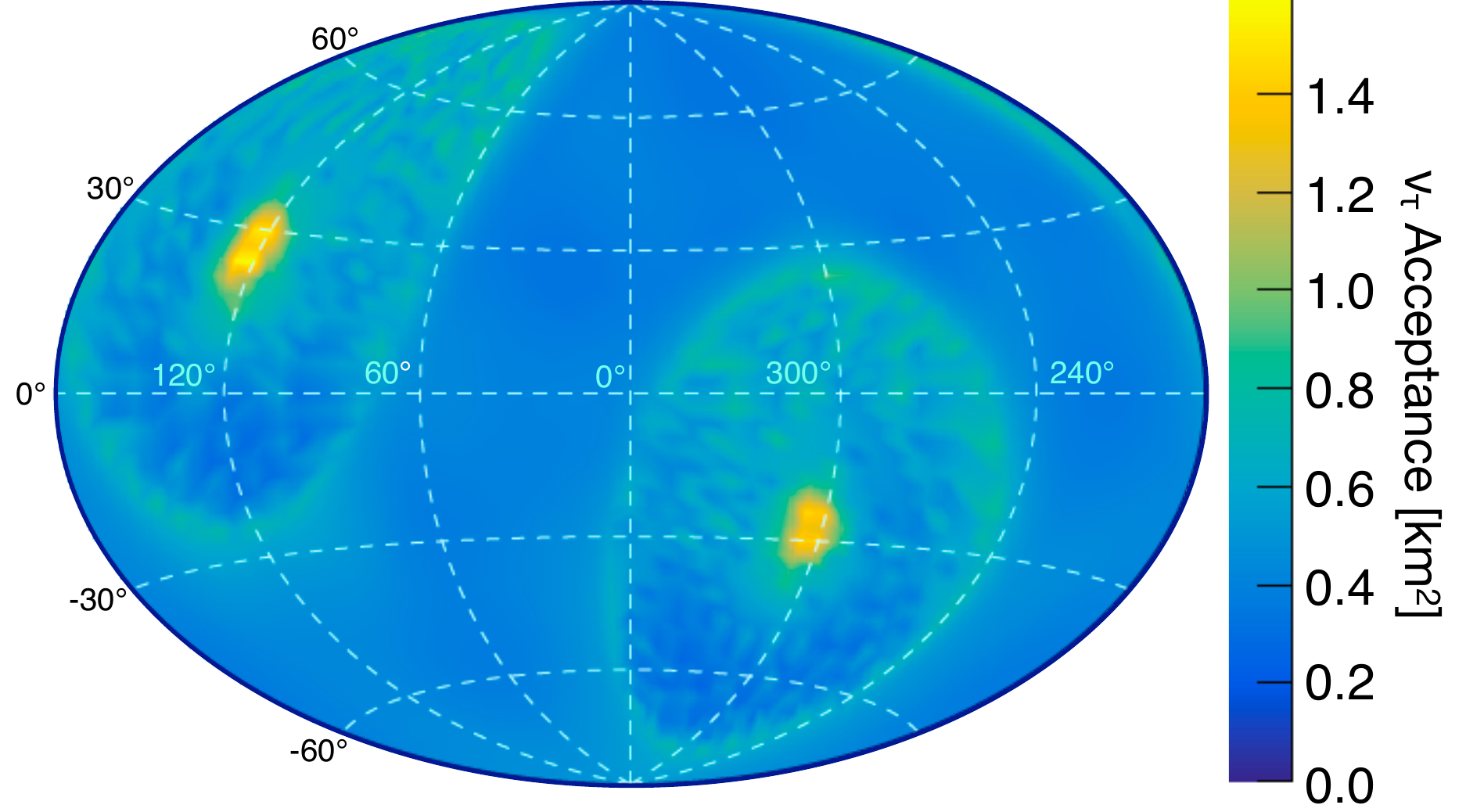}}
    \caption{\textit{{\rm Left:} All-flavor long-duration transient sensitivity band for POEMMA. For comparison, upper limits from IceCube for neutrino searches for GW170817~\cite{ANTARES:2017bia} and sensitivity projections for GRAND200k~\cite{Alvarez-Muniz:2018bhp} are also plotted. {\rm Right:} Sky plot of the time-averaged $\nu_{\tau}$ acceptance at $10^{8.5}$~GeV.}}\label{fig:POEMMA_Long}
    \vspace{-2.0ex}
\end{figure}

\begin{figure}
    \centering
    \raisebox{-0.5\height}{\includegraphics[width=0.35\textwidth, trim = 2.0in 1.0in 2.0in 1.0in, clip]{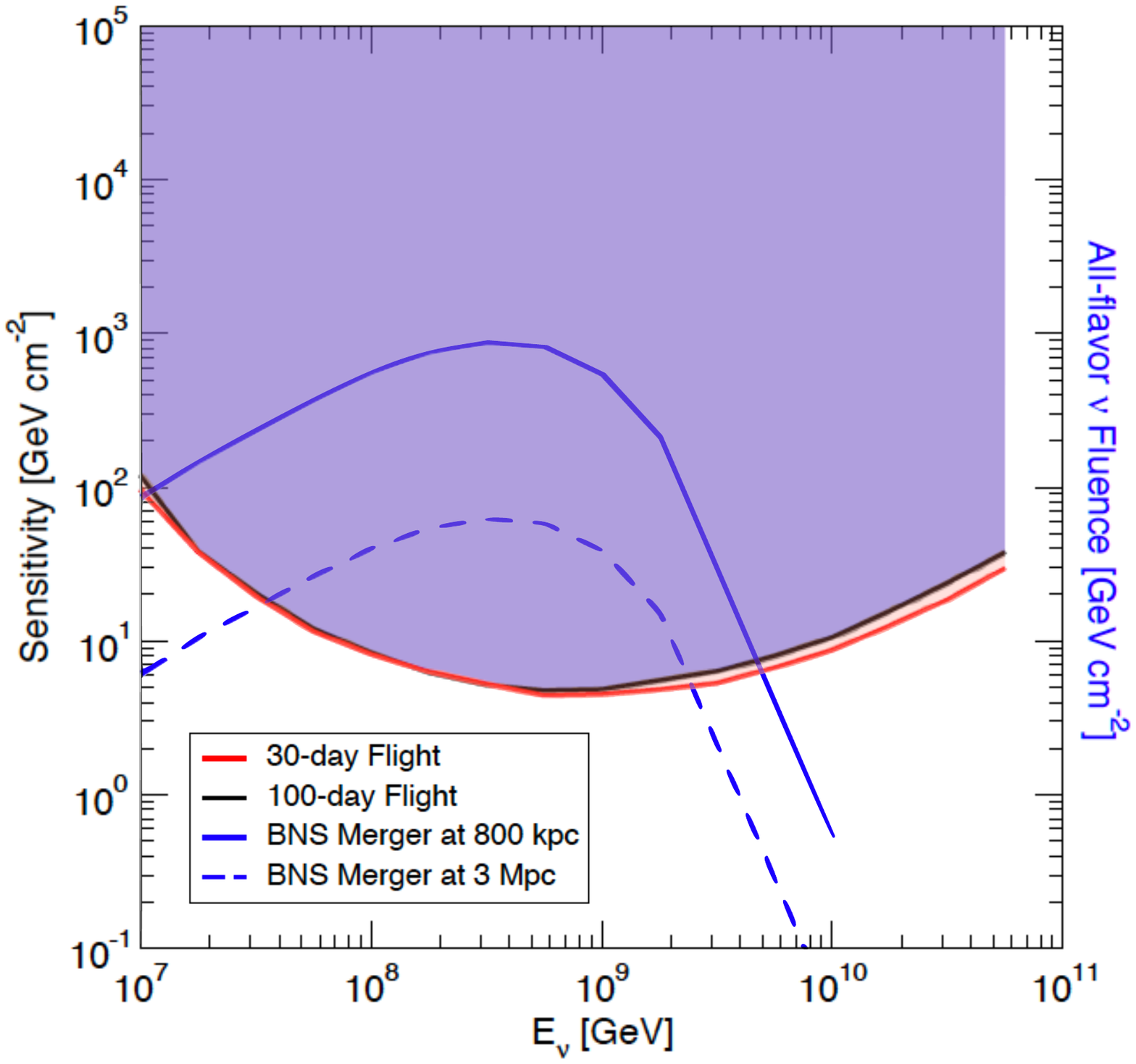}}\raisebox{-0.5\height}{\includegraphics[width=0.3\textwidth, trim = 1.75in 2.0in 1.75in 2.35in, clip]{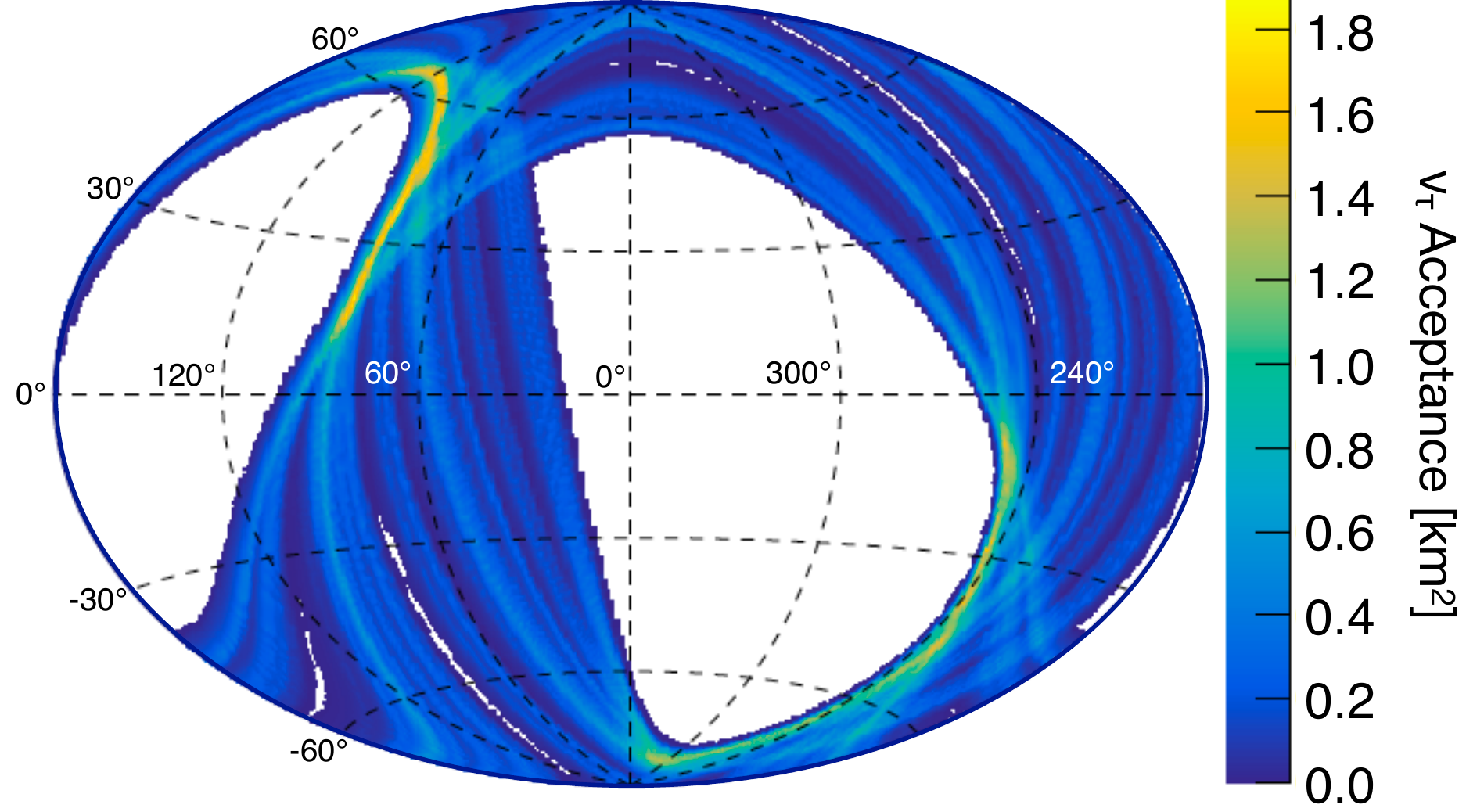}}\raisebox{-0.5\height}{\includegraphics[width=0.3\textwidth, trim = 1.75in 2.0in 1.75in 2.35in, clip]{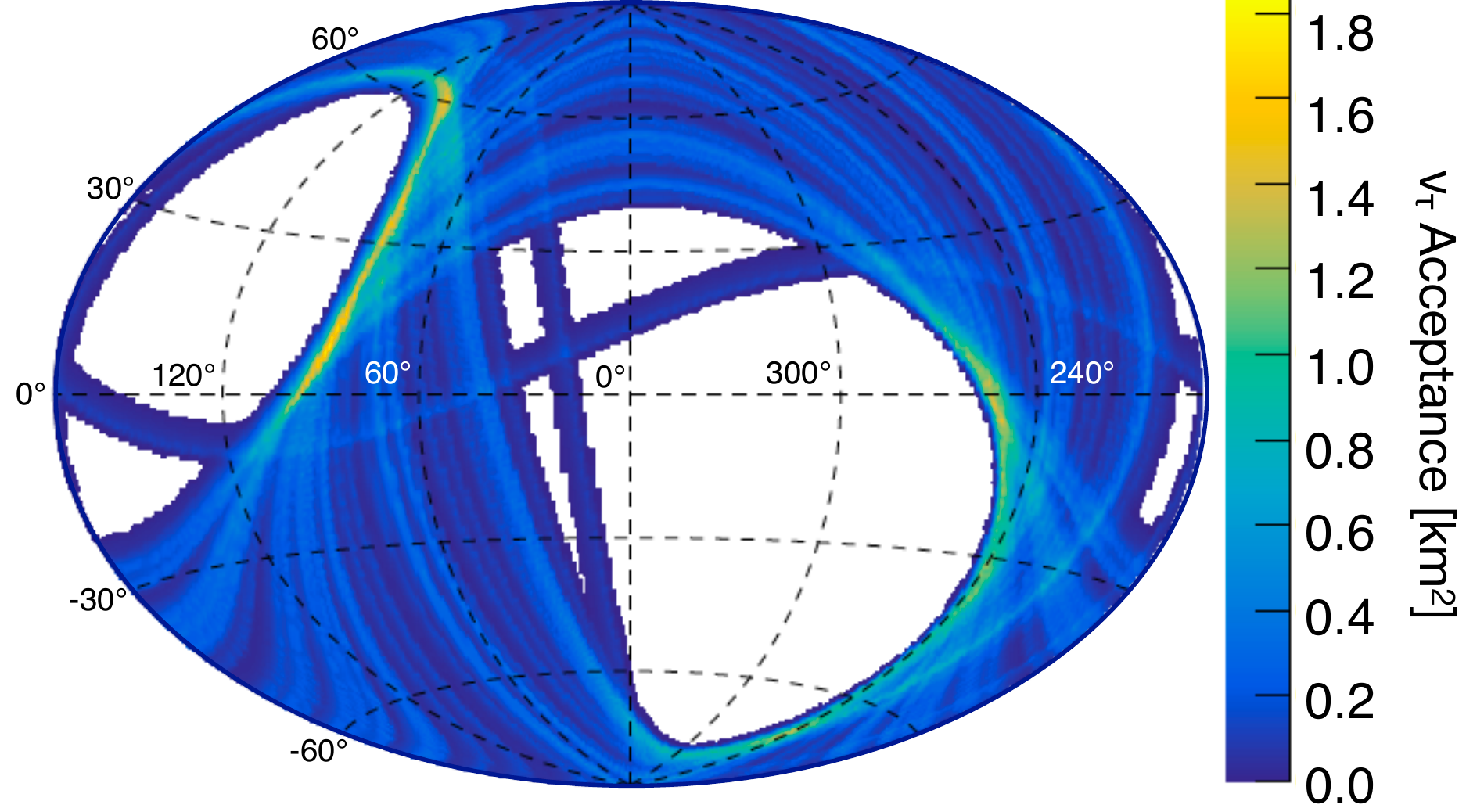}}
    \caption{\textit{{\rm Left}: All-flavor long-duration transient sensitivity band for EUSO-SPB2. {\rm Middle}: Sky plot of the $\nu_{\tau}$ acceptance at $10^{8.5}$~GeV for EUSO-SPB2 averaged over $30$~days. {\rm Right}: Sky plot of the $\nu_{\tau}$ acceptance at $10^{8.5}$~GeV for EUSO-SPB2 averaged over $100$~days.}}
    \label{fig:SPB2_Long}
    \vspace{-3.0ex}
\end{figure}

\begin{figure}
    \centering
    \raisebox{-0.5\height}{\includegraphics[width=.45\textwidth]{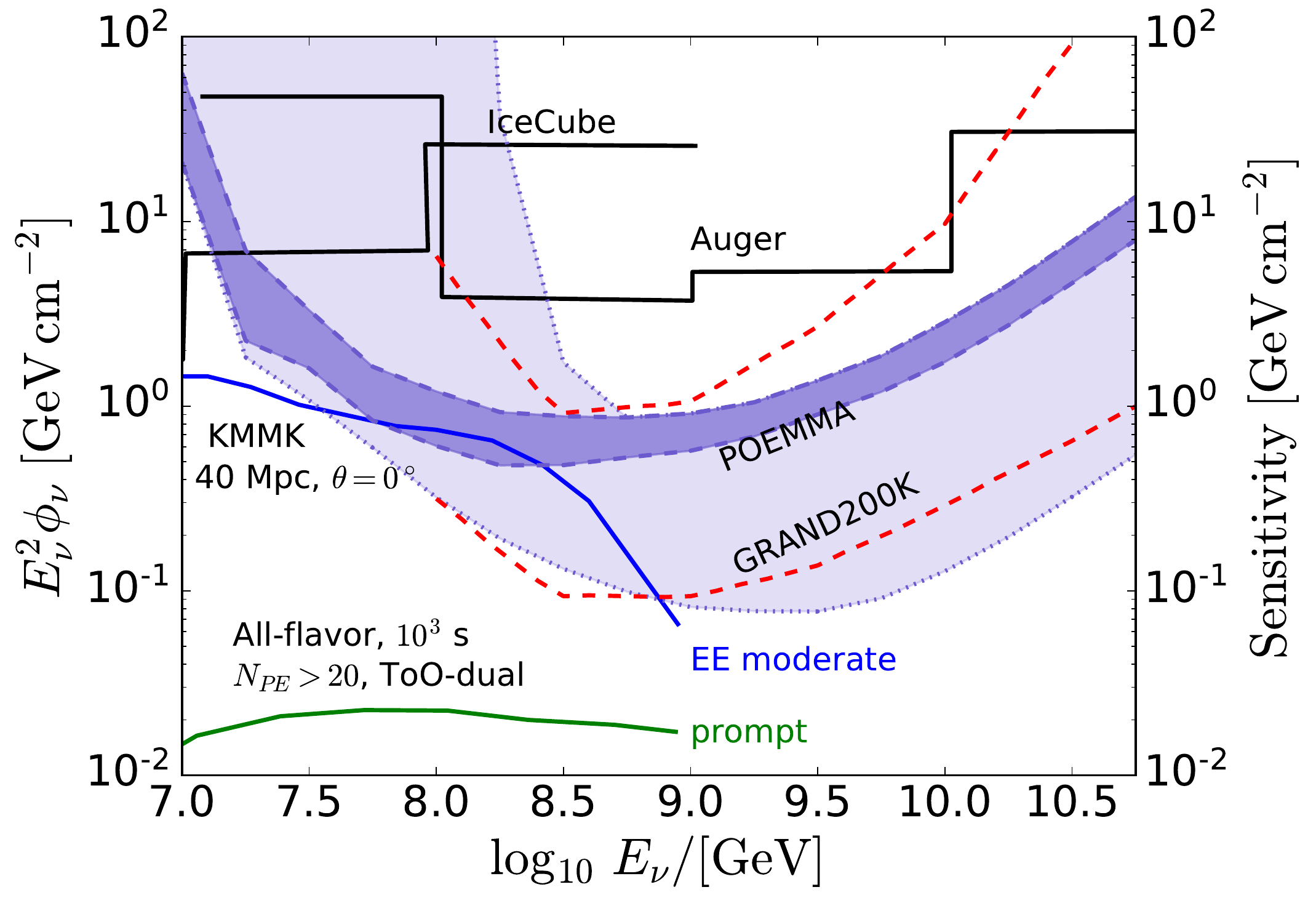}}\raisebox{-0.5\height}{\includegraphics[width=0.49\textwidth, trim = 1.75in 2.0in 1.75in 2.35in, clip]{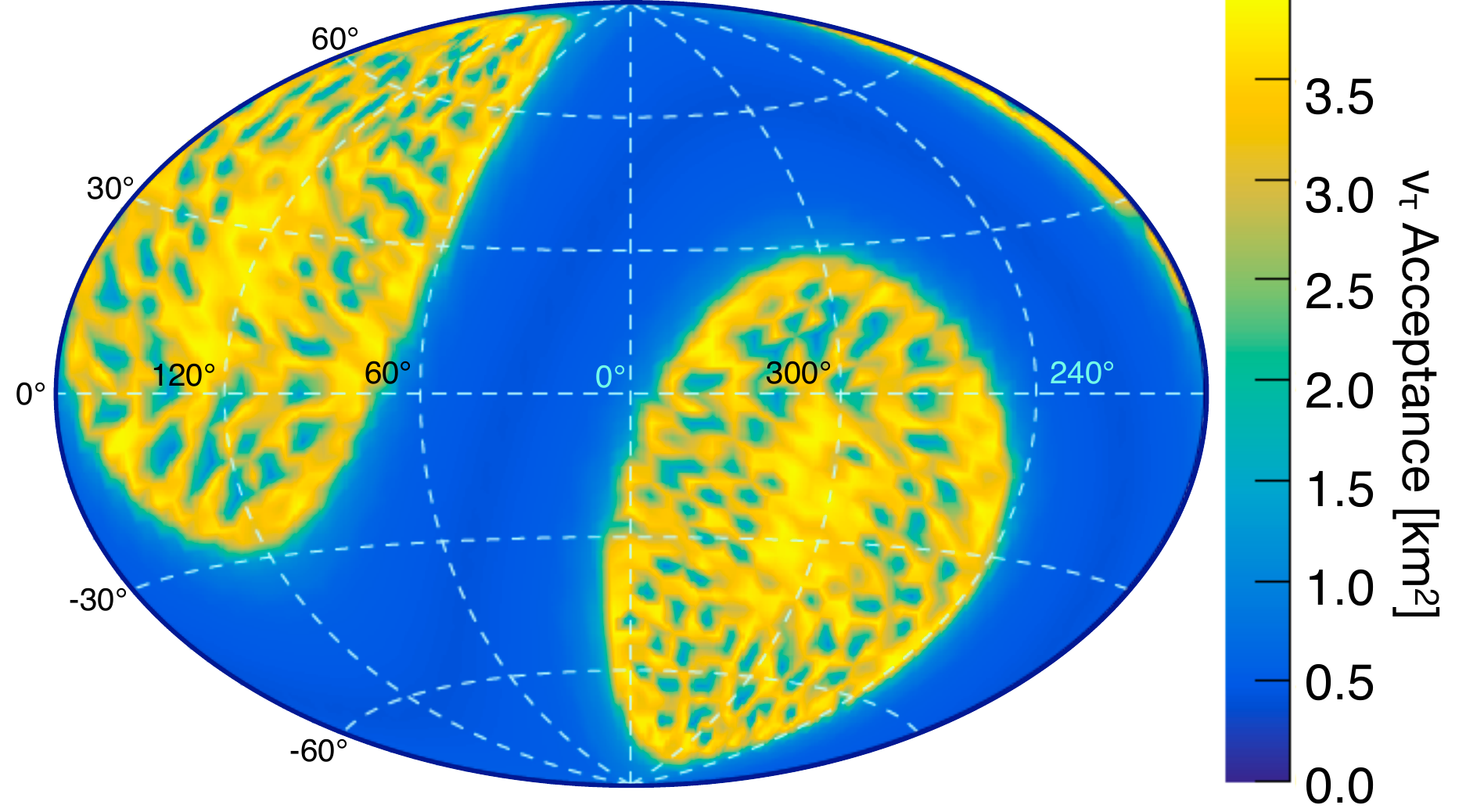}}
    \caption{\textit{{\rm Left}: Best-case all-flavor short-duration transient sensitivity band for POEMMA. For comparison, upper limits from IceCube and Auger for neutrino searches for GW170817~\cite{ANTARES:2017bia} and sensitivity projections for GRAND200k~\cite{Alvarez-Muniz:2018bhp} are also plotted. {\rm Right:} Sky plot of the time-averaged $\nu_{\tau}$ acceptance at $10^{8.5}$~GeV for POEMMA assuming the best-case short scenario.}}
    \label{fig:POEMMA_Short}
    \vspace{-2.0ex}
\end{figure}

\begin{figure}
    \centering
    \raisebox{-0.5\height}{\includegraphics[width=.45\textwidth, trim = 2.0in 1.0in 2.0in 1.0in, clip]{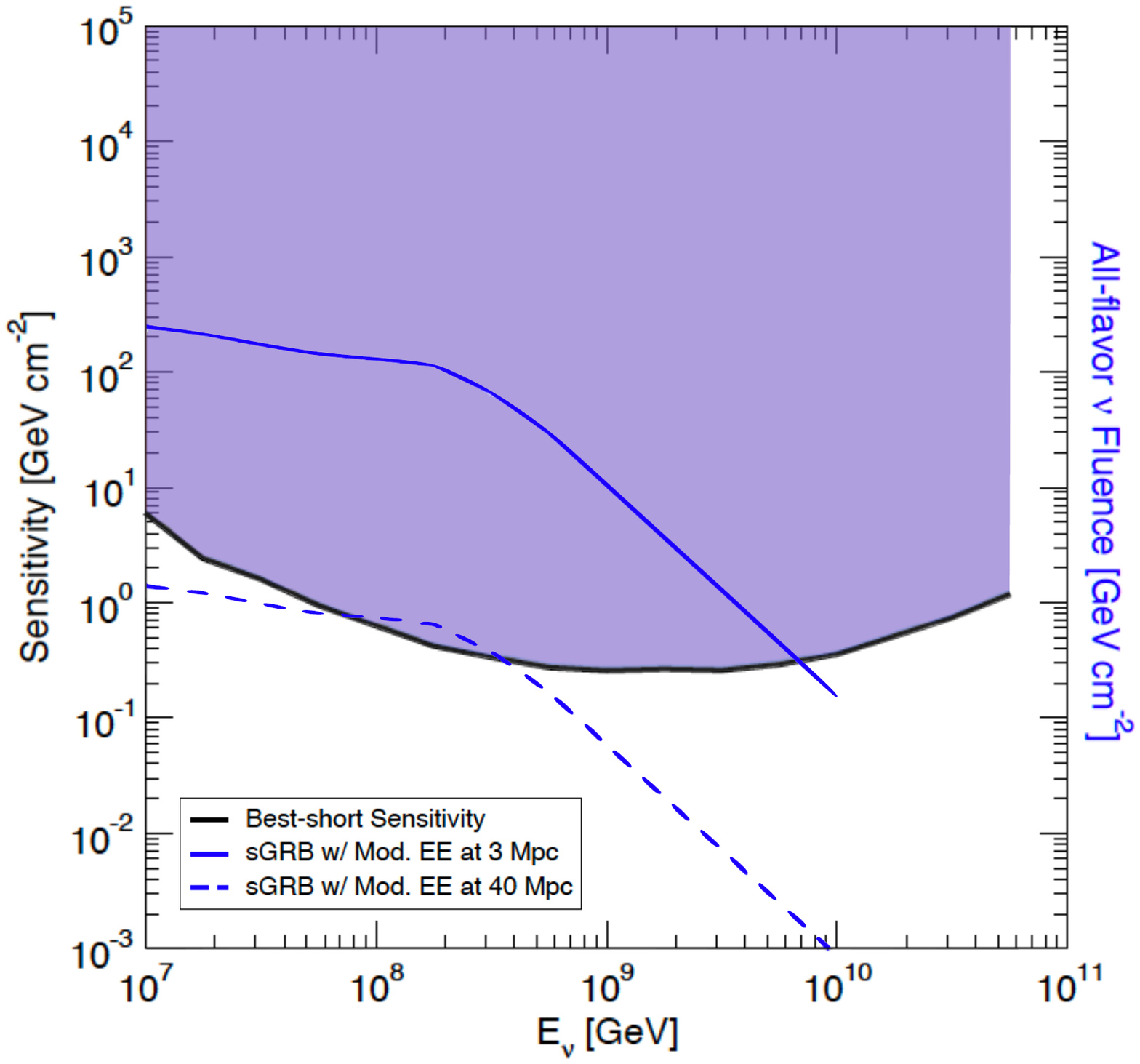}}\raisebox{-0.5\height}{\includegraphics[width=0.45\textwidth, trim = 1.75in 2.0in 1.75in 2.35in, clip]{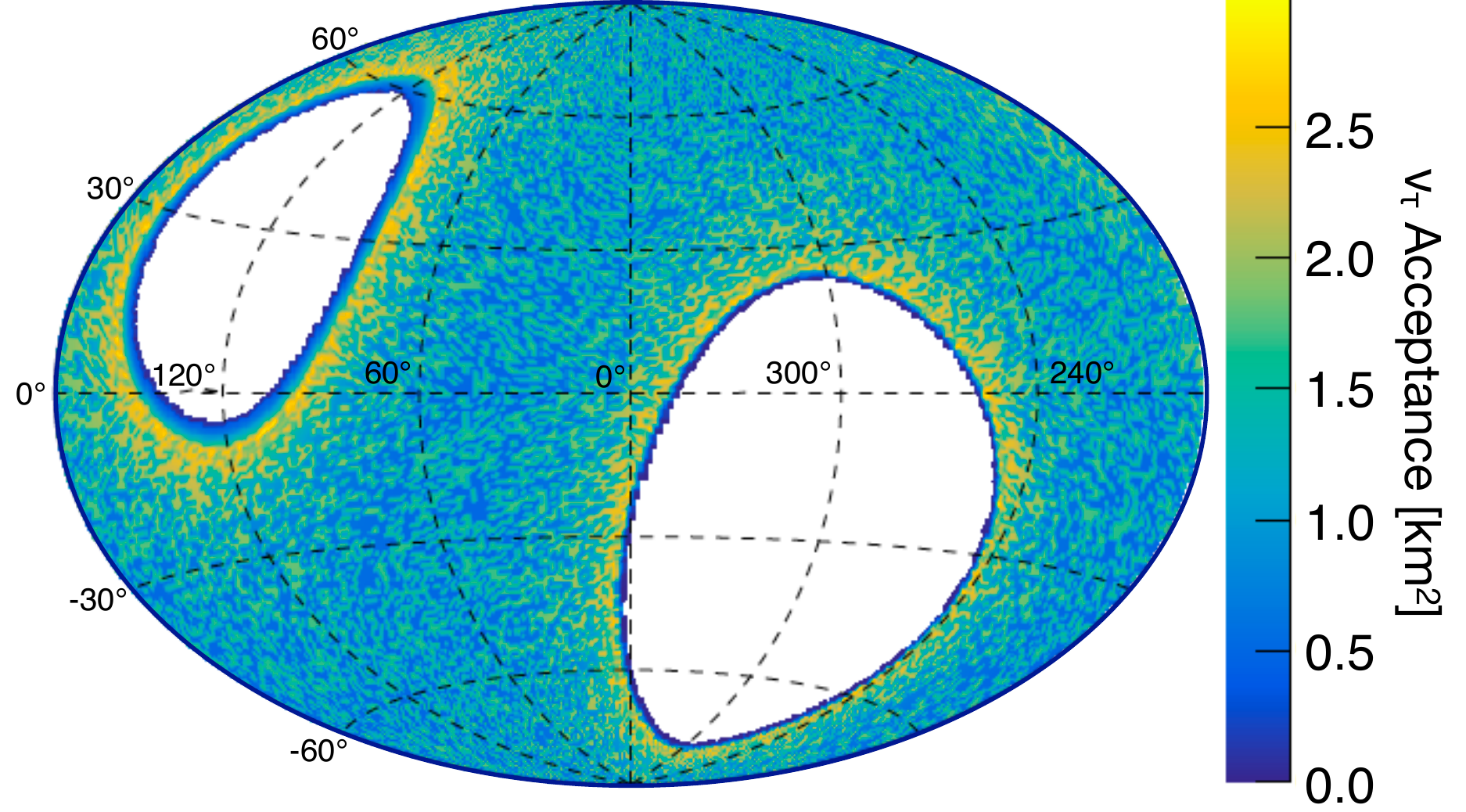}}
    \caption{\textit{{\rm Left:} Best-case all-flavor short-duration transient sensitivity band for EUSO-SPB2. {\rm Right:} Sky plot of the time-averaged $\nu_{\tau}$ acceptance at $10^{8.5}$~GeV for EUSO-SPB2 assuming the best-case short scenario.}}
    \label{fig:SPB2_Short}
    \vspace{-3.0ex}
\end{figure}

The left panels of Figures~\ref{fig:POEMMA_Long} and \ref{fig:SPB2_Long} show the long-duration all-flavor transient sensitivities for POEMMA and EUSO-SPB2, respectively. The sensitivities are presented as purple bands to demonstrate the variation across the sky. For comparison, modeled all-flavor neutrino fluences for a binary neutron star (BNS) merger~\cite{Fang_BNNMerger} at various time scales and distances are also plotted. Figures~\ref{fig:POEMMA_Long} and \ref{fig:SPB2_Long} also provide sky plots of the time-averaged $\nu_{\tau}$ acceptance at $10^{8.5}$~GeV to indicate the sky coverage for each instrument.
The left panels of Figures~\ref{fig:POEMMA_Short} and \ref{fig:SPB2_Short} show the best-case all-flavor transient sensitivities to short-duration events for POEMMA and EUSO-SPB2, respectively. For comparison, modeled all-flavor neutrino fluences for various phases of a short gamma-ray burst~\cite{Kimura:2017kan} at various distances are also plotted. Figures~\ref{fig:POEMMA_Short} and \ref{fig:SPB2_Short} also provide sky plots of the time-averaged $\nu_{\tau}$ acceptance at $10^{8.5}$~GeV assuming the best-case short-duration scenario. Figures~\ref{fig:POEMMA_Long}--\ref{fig:SPB2_Short} demonstrate that both POEMMA and EUSO-SPB2 will achieve transient sensitivities at the level of modeled neutrino fluences for nearby sources.

\vspace{-2.0ex}
\section{Prospects of Detecting Neutrinos for Candidate Transient Astrophysical Sources}
\vspace{-2.0ex}

In order to determine the probability for detecting at least one ToO from a given source class, we must calculate the neutrino horizon, defined as the redshift, $z_{\rm hor}$, at which a given instrument can expect to detect one $\nu_{\tau}$ event. The expected number of neutrino events from an astrophysical source at redshift $z$ is given by
\begin{equation}
    N_{\rm ev} = \int \phi_{\nu_{\tau}}\left(E_{\nu}, z\right) \left<A\left(E_{\nu}\right)\right> dE_{\nu}\,,
\end{equation}
where $\phi_{\nu_{\tau}}\left(E_{\nu}, z\right)$ is the observed single-flavor neutrino fluence~\cite{ToO_poemma}. With $z_{\rm hor}$, we can then find the cosmological volume in which POEMMA or EUSO-SPB2 would be able to detect tau neutrinos. Multiplying by the cosmological event rate for the source population taken from the literature gives the expected rate of ToOs, $r$, for the source class. Modeling the occurrence of transient events as a Poisson process, we can determine the probability of POEMMA or EUSO-SPB2 detecting at least one ToO as a function of time, $t$,
\begin{equation}
    P(\geq 1 \mbox{ToO}) = 1 - P(0) = 1 - e^{-rt}\,.
\end{equation}
Table~\ref{fig:HorTable} provides the neutrino horizon luminosity distances for POEMMA and EUSO-SPB2 for various neutrino source models. For these models, both POEMMA and EUSO-SPB2 will be able to detect neutrinos out to neighboring galaxies and beyond. The figure at right provides the Poisson probabilities of POEMMA detecting at least one ToO as a function of mission operation time for the various models. The most promising source classes for POEMMA are BNS mergers, binary black hole mergers, and tidal disruption events.

\begin{table}
\begin{minipage}[t]{0.58\textwidth}
\centering
\small
\begin{tabular}{|C{1.25in}|C{0.75in}|C{0.75in}|}
\hline
\hline
{Source Class} & {EUSO-SPB2 $\nu_{\tau}$ Horizon} & {POEMMA $\nu_{\tau}$ Horizon}\\ 
\hline
{TDE \ \ \ $M_{\rm BH} = 5 \times 10^{6} M_{\odot}$ }~\cite{Lunardini:2016xwi}& {$9$~Mpc} & {$128$~Mpc}\\ 
\hline
{TDE Base Scenario}~\cite{Lunardini:2016xwi} & {$4.5$~Mpc} & {$69$~Mpc} \\
\hline
{BBH merger -- Low Fluence}~\cite{Kotera:2016dmp} & {$6$~Mpc} & {$43$~Mpc} \\
\hline
{BBH merger -- High Fluence}~\cite{Kotera:2016dmp} & {$19$~Mpc} & {$137$~Mpc} \\
\hline
{BNS merger}~\cite{Fang_BNNMerger} & {$2.3$~Mpc} & {$16$~Mpc}\\ 
\hline
{sGRB w/ Mod. Ext. Emission}~\cite{Kimura:2017kan} & {$25$~Mpc} & {$90$~Mpc} \\
\hline
\hline
\end{tabular}{}
\end{minipage}
\begin{minipage}[t]{0.4\textwidth}
\raisebox{-0.5\height}{\includegraphics[trim = 2.75in 1.75in 2.75in 1.75in, clip, width = 1.0\textwidth]{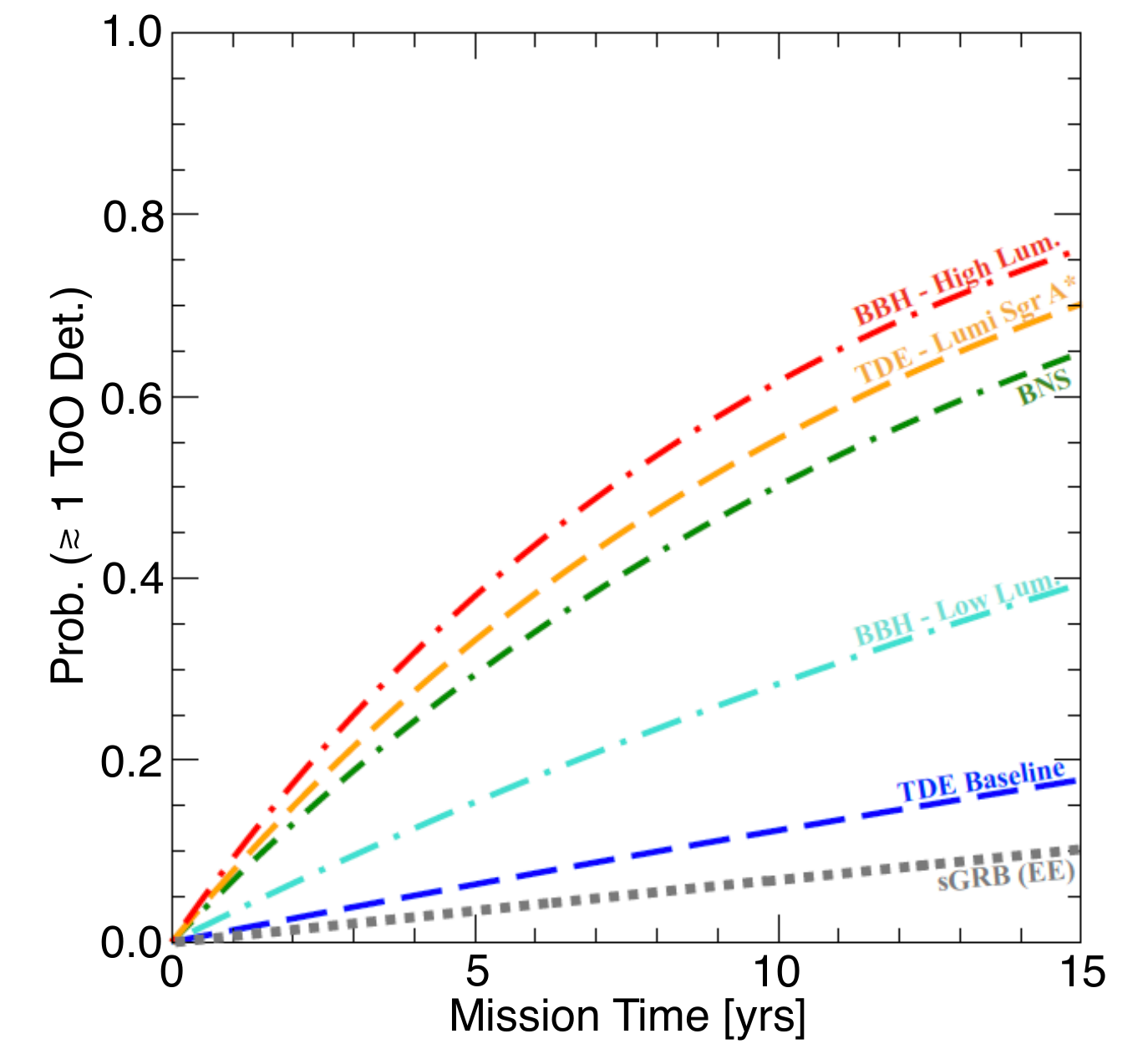}}
\end{minipage}
\caption{\textit{{\rm Left:} Table of horizon distances for EUSO-SPB2 and POEMMA for various astrophysical neutrino models. {\rm Right:} Poisson probability of POEMMA detecting at least one ToO versus mission time for several modeled source classes. Probabilities for EUSO-SPB2 are $\lesssim 1$\%. Plot reproduced from~\cite{ToO_poemma}.}} \label{fig:HorTable}
\vspace{-3.0ex}
\end{table}


\noindent \footnotesize{\textbf{\textit{Acknowledgements}}} -- This work is supported by NASA grants 80NSSC19K0626 at the University of Maryland, Baltimore County, 80NSSC19K0460 at the Colorado School of Mines, 80NSSC19K0484 at the University of Iowa, and 80NSSC19K0485 at the University of Utah, 80NSSC18K0464 at Lehman College, and under proposal 17-APRA17-0066 at NASA/GSFC and JPL. The conceptual design of POEMMA was supported by NASA Probe Mission Concept Study grant NNX17AJ82G for the 2020 Decadal Survey Planning. Contributors to this work were supported in part by NASA awards 16-APROBES16-0023, NNX17AJ82G, NNX13AH54G, 80NSSC18K0246, and 80NSSC18K0473.


\bibliography{ICRC2021ref.bib}

%

\clearpage
\section*{\Coll\ POEMMA Collaboration}
%
%
\scriptsize
\noindent
A. V. Olinto,$^1$
J. Krizmanic,$^{2,3}$
J. H. Adams,$^4$
R. Aloisio,$^5$
L. A. Anchordoqui,$^6$
A. Anzalone,$^{7,8}$
M. Bagheri,$^9$
D. Barghini,$^{10}$
M. Battisti,$^{10}$
D. R. Bergman,$^{11}$
M. E. Bertaina,$^{10}$
P. F. Bertone,$^{12}$
F. Bisconti,$^{13}$
M. Bustamante,$^{14}$
F. Cafagna,$^{15}$
R. Caruso,$^{16,8}$
M. Casolino,$^{17,18}$
K. \v{C}ern\'{y},$^{19}$
M. J. Christl,$^{12}$
A. L. Cummings,$^{5}$
I. De Mitri,$^{5}$
R. Diesing,$^{1}$
R. Engel,$^{20}$
J. Eser,$^{1}$
K. Fang,$^{21}$
F. Fenu,$^{10}$
G. Filippatos,$^{22}$
E. Gazda,$^{9}$
C. Guepin,$^{23}$
A. Haungs,$^{20}$
E. A. Hays,$^{2}$
E. G. Judd,$^{24}$
P. Klimov,$^{25}$
V. Kungel,$^{22}$
E. Kuznetsov,$^{4}$
\v{S}. Mackovjak,$^{26}$
D. Mand\'{a}t,$^{27}$ 
L. Marcelli,$^{18}$ 
J. McEnery,$^{2}$ 
G. Medina-Tanco,$^{28}$ 
K.-D. Merenda,$^{22}$ 
S. S. Meyer,$^{1}$
J. W. Mitchell,$^{2}$ 
H. Miyamoto,$^{10}$ 
J. M. Nachtman,$^{29}$
A. Neronov,$^{30}$ 
F. Oikonomou,$^{31}$ 
Y. Onel,$^{29}$ 
G. Osteria,$^{32}$ 
A. N. Otte,$^{9}$ 
E. Parizot,$^{33}$ 
T. Paul,$^{6}$ 
M. Pech,$^{27}$ 
J. S. Perkins,$^{2}$
P. Picozza,$^{18,34}$ 
L.W. Piotrowski,$^{35}$ 
Z. Plebaniak,$^{10}$ 
G. Pr\'{e}v\^{o}t,$^{33}$ 
P. Reardon,$^{4}$
M. H. Reno,$^{29}$ 
M. Ricci,$^{36}$ 
O. Romero Matamala,$^{9}$
F. Sarazin,$^{22}$ 
P. Schov\'{a}nek,$^{27}$ 
V. Scotti,$^{32,37}$
K. Shinozaki,$^{38}$ 
J. F. Soriano,$^{6}$ 
F. Stecker,$^{2}$
Y. Takizawa,$^{17}$ 
R. Ulrich,$^{20}$ 
M. Unger,$^{20}$ 
T. M. Venters,$^{2}$ 
L. Wiencke,$^{22}$ 
D. Winn,$^{29}$ 
R. M. Young,$^{12}$
M. Zotov$^2$\\

%

\noindent
$^{1}$The University of Chicago, Chicago, IL, USA
$^{2}$NASA Goddard Space Flight Center, Greenbelt, MD, USA
$^{3}$Center for Space Science \& Technology, University of Maryland, Baltimore County, Baltimore, MD, USA
$^{4}$University of Alabama in Huntsville, Huntsville, AL, USA
$^{5}$Gran Sasso Science Institute, L'Aquila, Italy
$^{6}$City University of New York, Lehman College, NY, USA
$^{7}$Istituto Nazionale di Astrofisica INAF-IASF, Palermo, Italy
$^{8}$Istituto Nazionale di Fisica Nucleare, Catania, Italy
$^{9}$Georgia Institute of Technology, Atlanta, GA, USA
$^{10}$Universita' di Torino, Torino, Italy
$^{11}$University of Utah, Salt Lake City, Utah, USA
$^{12}$NASA Marshall Space Flight Center, Huntsville, AL, USA
$^{13}$Istituto Nazionale di Fisica Nucleare, Turin, Italy
$^{14}$Niels Bohr Institute, University of Copenhagen, DK-2100 Copenhagen, Denmark
$^{15}$Istituto Nazionale di Fisica Nucleare, Bari, Italy
$^{16}$Universita' di Catania, Catania Italy
$^{17}$RIKEN, Wako, Japan
$^{18}$Istituto Nazionale di Fisica Nucleare, Section of Roma Tor Vergata, Italy
$^{19}$Joint Laboratory of Optics,  Faculty of Science,  Palack\'{y} University,  Olomouc,  Czech Republic
$^{20}$Karlsruhe Institute of Technology, Karlsruhe, Germany
$^{21}$Kavli Institute for Particle Astrophysics and Cosmology, Stanford University, Stanford, CA94305, USA
$^{22}$Colorado School of Mines, Golden, CO, USA
$^{23}$Department of Astronomy, University of Maryland, College Park, MD, USA
$^{24}$Space Sciences Laboratory, University of California, Berkeley, CA, USA
$^{25}$Skobeltsyn Institute of Nuclear Physics, Lomonosov Moscow State University, Moscow,Russia
$^{26}$Institute of Experimental Physics, Slovak Academy of Sciences, Kosice, Slovakia
$^{27}$Institute of Physics of the Czech Academy of Sciences, Prague, Czech Republic
$^{28}$Instituto de Ciencias Nucleares, UNAM, CDMX, Mexico
$^{29}$University of Iowa, Iowa City, IA, USA
$^{30}$University of Geneva, Geneva, Switzerland
$^{31}$Institutt for fysikk, NTNU, Trondheim, Norway
$^{32}$Istituto Nazionale di Fisica Nucleare, Napoli, Italy
$^{33}$Universit\'{e} de Paris, CNRS, Astroparticule et Cosmologie, F-75013 Paris, France
$^{34}$Universita di Roma Tor Vergata, Italy
$^{35}$Faculty of Physics, University of Warsaw, Warsaw, Poland
$^{36}$Istituto Nazionale di Fisica Nucleare - Laboratori Nazionali di Frascati, Frascati, Italy
$^{37}$Universita' di Napoli Federico II, Napoli, Italy
$^{38}$National Centre for Nuclear Research, Lodz, Poland
\section*{\Coll\ JEM-EUSO Collaboration}
\scriptsize
\noindent
\begin{sloppypar}
{\scriptsize \noindent 
G.~Abdellaoui$^{ah}$, 
S.~Abe$^{fq}$, 
J.H.~Adams Jr.$^{pd}$, 
D.~Allard$^{cb}$, 
G.~Alonso$^{md}$, 
L.~Anchordoqui$^{pe}$,
A.~Anzalone$^{eh,ed}$, 
E.~Arnone$^{ek,el}$,
K.~Asano$^{fe}$,
R.~Attallah$^{ac}$
% ...
}
\end{sloppypar}
\vspace*{.3cm}

\begin{sloppypar}{ \scriptsize
\noindent
% ...
$^{ac}$LPR at Dept. of Physics, Faculty of Sciences, University Badji Mokhtar, Annaba, Algeria,
% ...
$^{ah}$Telecom Lab., Faculty of Technology, University Abou Bekr Belkaid, Tlemcen, Algeria,
% ...
$^{cb}$Universit\'e de Paris, CNRS, AstroParticule et Cosmologie, F-75013 Paris, France,
% ...
$^{ed}$Istituto Nazionale di Fisica Nucleare - Sezione di Catania, Italy,
% ...
$^{eh}$INAF - Istituto di Astrofisica Spaziale e Fisica Cosmica di Palermo, Italy,
% ...
$^{ek}$Istituto Nazionale di Fisica Nucleare - Sezione di Torino, Italy,
$^{el}$Dipartimento di Fisica, Universita' di Torino, Italy,
% ...
$^{fe}$Institute for Cosmic Ray Research, University of Tokyo, Kashiwa, Japan,
% ...
$^{fq}$Nihon University Chiyoda, Tokyo, Japan, 
% ...
$^{md}$Universidad Polit\'ecnia de madrid (UPM), Madrid, Spain,
% ...
$^{pd}$University of Alabama in Huntsville, Huntsville, AL, USA,
$^{pe}$Lehman College, City University of New York (CUNY),NY, USA,
% ...
}
\end{sloppypar}
\vspace*{.3cm}

\end{document}